# High-$T_c$ superconductivity up to 55 K under high pressure in the heavily electron doped Li$_x$(NH$_3$)$_y$Fe$_2$Se$_2$ single crystal

P. Shahi,[1,2=] J. P. Sun,[1,2=] S. S. Sun,[3=] Y. Y. Jiao,[1,2] K. Y. Chen,[1,2] S. H. Wang,[3] H. C. Lei,[3*] Y. Uwatoko,[4] B. S. Wang,[1,2] and J.-G. Cheng[1,2*]

[1]Beijing National Laboratory for Condensed Matter Physics and Institute of Physics, Chinese Academy of Sciences, Beijing 100190, China

[2] School of Physical Sciences, University of Chinese Academy of Sciences, Beijing 100190, China

[3]Department of Physics and Beijing Key Laboratory of Opto-electronic Functional Materials & Micro-nano Devices, Renmin University of China, Beijing 100872, China

[4]Institute for Solid State Physics, University of Tokyo, 5-1-5 Kashiwanoha, Kashiwa, Chiba 277-8581, Japan

= These authors contributed to this work.

E-mails: jgcheng@iphy.ac.cn, or hlei@ruc.edu.cn

## Abstract

We report a high-pressure study on the heavily electron doped Li$_x$(NH$_3$)$_y$Fe$_2$Se$_2$ single crystal by using the cubic anvil cell apparatus. The superconducting transition temperature $T_c \approx 44$ K at ambient pressure is first suppressed to below 20 K upon increasing pressure to $P_c \approx 2$ GPa, above which the pressure dependence of $T_c(P)$ reverses and $T_c$ increases steadily to ca. 55 K at 11 GPa. These results thus evidenced a pressure-induced second high-$T_c$ superconducting (SC-II) phase in Li$_x$(NH$_3$)$_y$Fe$_2$Se$_2$ with the highest $T_c^{max} \approx 55$K among the FeSe-based ***bulk*** materials. Hall data confirm that in the emergent SC-II phase the dominant electron-type carrier density undergoes a fourfold enhancement and tracks the same trend as $T_c(P)$. Interesting, we find a nearly parallel scaling behavior between $T_c$ and the inverse Hall coefficient for the SC-II phases of both Li$_x$(NH$_3$)$_y$Fe$_2$Se$_2$ and (Li,Fe)OHFeSe. The present work demonstrates that high pressure offers a distinctive means to further raising the maximum $T_c$ of heavily electron doped FeSe-based materials by increasing the effective charge carrier concentration via a plausible Fermi surface reconstruction at $P_c$.

To find out the approaches to raise the critical temperature $T_c$ of unconventional superconductors is one of the most enduring problems in the contemporary condensed matter physics. The structural simplest FeSe and its derived materials offer a fertile ground to address this issue. In particular, the great tunability of $T_c$ for bulk FeSe from 8 K to over 40 K [1-7] and the possible high $T_c$ exceeding the liquid-nitrogen temperature in the monolayer FeSe/SrTiO$_3$ [8,9] have spurred tremendous research interest recently. The principal route to raise the $T_c$ of FeSe is to doping electron, which has been successfully achieved via the interlayer intercalations [$A_x$Fe$_{2-y}$Se$_2$ (A = K, Rb, …), $A_x$(NH$_3$)$_y$Fe$_2$Se$_2$, and (Li,Fe)OHFeSe] [2,3,7,10,11], interface charge transfer (monolayer FeSe/SrTiO$_3$) [8], surface K dosing [12,13], and gate-voltage regulation [6,14]. A common Fermi surface topology consisting of electron pockets only have been confirmed by the APRES measurements on these heavily electron-doped (HED) FeSe derivatives [15-17].

Based on the bulk resistivity measurements on these HED FeSe-derived materials, the highest $T_c$ at ambient pressure, *i.e.* $T_c^{onset}$ = 46.6 K and $T_c^{zero}$ = 44.8 K, was achieved in the FeSe flake under a regulated voltage of 4.5 V in a field-effect transistor device based on a solid ion conductor [14]. Although the electron doping has been widely believed to play an essential role to raise the $T_c$, further enhancement of its $T_c$ via adding more electrons seems to be plagued by the observed insulating state in the overdoped regime as seen in the gate-voltage regulation and surface K dosing experiments on FeSe and (Li,Fe)OHFeSe [14,18,19]. Whether the $T_c$ of the HED FeSe-based bulk materials can reach well above 50 K or even approach to that of monolayer FeSe/SrTiO$_3$ remains an open issue.

Given the limitations of electron doping, it is imperative to explore other routes to further enhance the $T_c$ of these HED FeSe-based materials. The application of high pressure can provide an alternative means. It has been reported that pressurization on some HED FeSe-based materials, *e.g.* $A_x$Fe$_{2-y}$Se$_2$ (A = K, Rb, …) [20,21] and Cs$_{0.4}$(NH$_3$)$_y$FeSe [22], can first reduce $T_c$ and then above a critical pressure $P_c$ induce a second high-$T_c$ superconducting phase (denoted as SC-II to distinguish from the ambient pressure SC-I phase). The observed $T_c$ in the SC-II phase is ca. 10 K higher than that of SC-I phase. Our recent high-pressure study on the (Li,Fe)OHFeSe single crystals also evidenced such an SC-II phase above $P_c \approx 5$ GPa, reaching a record high $T_c^{onset}$ = 51 K and $T_c^{zero}$ = 46.5 K at 12.5 GPa [23]. This is the first time to raise the $T_c$ of FeSe-based **bulk** materials (excluding the monolayer FeSe) to above 50 K, raising the hope for further enhancement of $T_c$. More intriguingly, we observed a sharp transition of the normal-state properties of (Li,Fe)OHFeSe from a Fermi-liquid for SC-I to non-Fermi-liquid for SC-II phase. In addition, the emergence of SC-II phase is accompanied by a concurrent enhancement of electron carrier density. These observations demonstrated that high pressure plays a very distinctive role to tune the superconducting properties of these HED FeSe-based materials.

In order to avoid the complications from the magnetism of intercalated (Li,Fe)OH layers in (Li,Fe)OHFeSe [3], we turn our attention to the recently synthesized $Li_x(NH_3)_yFe_2Se_2$ single crystals [24], which can reach an optimal $T_c^{onset} \approx 44.3$ K at the ambient-pressure SC-I phase. By performing high-pressure magneto-transport measurements up to 12 GPa, we find that the SC-I phase is quickly suppressed under a low $P_c \approx 2$ GPa, above which a high-$T_c$ SC-II phase emerges and the highest $T_c^{onset}$ reaches ca. 55 K above 10 GPa. Similar with (Li,Fe)OHFeSe [23], the reemergence of the SC-II phase is also accompanied with a concurrent enhancement of the electron carrier density. Importantly, we obtain a linear relationship between $T_c$ and the inverse Hall coefficient for the SC-II phases in both systems. Our present work thus demonstrates a unique way of high pressure to further raising $T_c$ of these HED FeSe-based materials via increasing the effective electron carrier density.

Details about the crystal growth and characterizations of $Li_x(NH_3)_yFe_2Se_2$ single crystals at ambient pressure can be found elsewhere [24]. A palm cubic anvil cell (CAC) apparatus was employed for the accurate measurements of magneto-transport and ac magnetic susceptibility under hydrostatic pressures up to 12 GPa [25]. The standard four-probe method was used for the resistivity measurement, and the current is applied within the *ab* plane with the magnetic field along the *c* axis. Anti-symmetrized (symmetrized) method was performed to get the $\rho_{xy}(H)$ and $\rho_{xx}(H)$ data. The mutual induction method was employed for the ac magnetic susceptibility measurements with an excitation current of 1 mA and 317 Hz. The superconducting shielding volume fraction was estimated by comparing with the superconducting signal of Pb. The glycerol was used as the pressure transmitting medium and the pressure values inside the CAC were calibrated at room temperature by measuring the characteristic transitions of bismuth and lead from resistivity.

Fig. 1(a) shows the temperature dependence of resistivity $\rho(T)$ under various hydrostatic pressures up to 12 GPa in the whole temperature range for the $Li_x(NH_3)_yFe_2Se_2$ single crystal with an optimal $T_c$ at ambient pressure. Here, we determine $T_c^{zero}$ (down-pointing arrow) as the zero-resistivity temperature, and define the onset $T_c^{onset}$ (up-pointing arrow) as the temperature where $\rho(T)$ curves above and below intersect with each other. At ambient pressure, the $\rho(T)$ curve displays a broad hump centered around 220 K and shows a sharp superconducting transition with $T_c^{onset}$ = 44.3 K and $T_c^{zero}$ = 42 K, in agreement with the previous report [24]. The normal-state $\rho(T)$ decreases considerably and the hump feature fades away gradually with increasing pressure to 3 GPa, above which a quasi-linear behavior is restored in the large temperature range. Similar behavior has also been observed in (Li,Fe)OHFeSe [23]. As illustrated in Fig. 1(a), the $\rho(T)$ curves at $P$ > 10 GPa exhibit an anomalous bendover feature in the normal state. As will be discussed below, such a behavior should be attributed to the partial conversion of $Li_x(NH_3)_yFe_2Se_2$ to the pristine FeSe, which then transforms to the three-dimensional MnP-type structure with a semiconducting behavior above 10 GPa [26].

At low temperatures, the superconducting transition displays a non-monotonic variation with pressure, which can be seen more clearly in Fig. 1(b) from the vertically shifted $\rho(T)$ curves below 100 K. As can be seen, the application of high pressure first reduces $T_c$ quickly to $T_c^{onset}$ = 25 K and $T_c^{zero}$ = 15 K at 2 GPa. Interestingly, the pressure dependence of $T_c(P)$ suddenly reverses at $P$ > 2 GPa and $T_c^{onset}$ ($T_c^{zero}$) increases to 37 K (26 K) at 3 GPa, thus *evidencing the emergence of SC-II phase* as seen in (Li,Fe)OHFeSe [23]. But the critical pressure $P_c \approx 2$ GPa is lower than that of (Li,Fe)OHFeSe. As illustrated by the arrows in Fig. 1(b), the $T_c^{onset}$ and $T_c^{zero}$ exhibit distinct pressure dependences at $P$ > 2 GPa: $T_c^{onset}$ first increases quickly with pressure to ~ 50 K at 6 GPa and then tends to level off, reaching the highest 55 K at 11 GPa, and finally decreases slightly with pressure; in contrast, $T_c^{zero}$ first tracks the $T_c^{onset}$ and reaches the maximum value of ~ 40 K at 6 GPa, and then decreases quickly with the difference between $T_c^{onset}$ and $T_c^{zero}$ enlarged considerably at $P$ > 6 GPa. Eventually, $T_c^{zero}$ can be barely reached at 11 and 12 GPa despite of a very high $T_c^{onset} \approx$ 55 K. Since our previous studies have proved an excellent hydrostatic pressure condition up to 15 GPa for CAC [25], the observed large discrepancy between $T_c^{onset}$ and $T_c^{zero}$ above 6 GPa reflects an intrinsic pressure response of Li$_x$(NH$_3$)$_y$Fe$_2$Se$_2$, implying that the superconducting transition either consists of a distribution of different $T_c$s or is not bulk in nature.

To further track the evolutions of $T_c(P)$ and to investigate the nature of the broad superconducting transitions above 6 GPa, we also measured the ac magnetic susceptibility $4\pi\chi(T)$ up to 11 GPa. As shown in Fig. 2(a) for $P$ < 6 GPa, a single superconducting diamagnetic drop can be clearly observed below $T_c^\chi$, which increases with pressure in perfect agreement with the $\rho(T)$ data. In addition, the superconducting shielding volume fraction increases with pressure, and reaches ~ 90% at 5 GPa, signaling a bulk nature for the observed SC-II phase at $P \leq 6$ GPa. In contrast, the $4\pi\chi(T)$ curves at 7 and 9 GPa, Fig. 2(b), evidence two superconducting transitions as indicated by the two successive drops, which correspond well to the $T_c^{onset}$ and $T_c^{zero}$ determined from $\rho(T)$ curves. The superconducting phase with a higher $T_c$ ~ 50 K can attain a volume fraction of ca. 30% and induces the sudden drop of resistivity at $T_c^{onset}$, but the sample can reach zero resistivity only when the lower-$T_c$ phase also enters the superconducting state below ~33 K. While the higher $T_c$ phase keeps nearly unchanged with pressure, both the $T_c$ and the volume fraction of the lower-$T_c$ phase decrease and nearly vanishes at 11 GPa. As such, the $\rho(T)$ at 11 GPa can hardly reach zero resistivity until very low temperatures.

From these above characterizations, we can conclude that the SC-II phase is bulk in nature for $P \leq 6$ GPa, whereas the sample contains two superconducting phases with different $T_c$s above 6 GPa: the high-$T_c$ ($\geq 50$ K) phase has a small but nearly constant volume fraction ~ 30% to 11 GPa, whereas the low-$T_c$ ($\leq 33$) phase shrinks and vanishes completely above 11 GPa. Fig. 3 summarizes the pressure dependences of these characteristic temperatures $T_c^{onset}$, $T_c^{zero}$, and $T_c^\chi$ for Li$_x$(NH$_3$)$_y$Fe$_2$Se$_2$ together with the $T_c^{zero}$ of FeSe single crystal for comparison [26]. Since the obtained $T_c^\chi$ of the low-$T_c$ phase and the $T_c^{zero}$ from $\rho(T)$ match perfectly with the $T_c^{zero}$ of FeSe single crystal at $P \geq 6$ GPa, it is very likely that Li$_x$(NH$_3$)$_y$Fe$_2$Se$_2$ has been partially converted back to the pristine FeSe due to the extrusion of Li$^+$ and ammonia under elevated pressures. Such a speculation is supported by the disappearance of the low-$T_c$ superconducting phase at 11 GPa when the pristine layered FeSe will transform to the three-dimensional MnP-type structure with a semiconducting behavior according to our previous study [26]. The presence of semiconducting FeSe above 10 GPa can also explain the observed bendover feature in the $\rho(T)$ curves of 11 and 12 GPa shown in Fig. 1(a). Nevertheless, the smooth evolution of the $T_c^{onset}$ and $T_c^\chi$ of the high-

$T_c$ phase at $P > 6$ GPa should reflect the intrinsic pressure responses of the remaining pressure-induced SC-II phase. Below we thus focus on the variations of $T_c^{onset}$ and $T_c^\chi$ as function of pressure for $Li_x(NH_3)_yFe_2Se_2$.

The temperature-pressure phase diagram shown in Fig. 3 depicts explicitly the evolution of the superconducting phases of $Li_x(NH_3)_yFe_2Se_2$ under pressure. The high-$T_c$ SC-I phase, initially achieved at ambient pressure via electron doping through inserting $Li^+$ and ammonia in between the FeSe layers, is quickly suppressed by pressure, and the SC-II phase emerges above $P_c \approx 2$ GPa and exists in a large pressure range. The pressure-induced SC-II phase in the $Li_x(NH_3)_yFe_2Se_2$ resembles those observed in $A_xFe_{2-y}Se_2$ [20,21], $Cs_{0.4}(NH_3)_yFeSe$ [22], and $(Li_{1-x}Fe_x)OHFeSe$ [23], pointing to a universal phenomenon for these HED FeSe-based materials under high pressure. But, some specific features for the present $Li_x(NH_3)_yFe_2Se_2$ are noteworthy; *i.e.* the critical pressure $P_c \approx 2$ GPa for the emergence of SC-II phase is the lowest while the maximum $T_c^{onset} \approx 55$ K is the highest among the studied HED FeSe-derived materials. For comparison, the $P_c$ and the maximum $T_c^{onset}$ are 5 GPa, 51 K for $(Li_{1-x}Fe_x)OHFeSe$ [23], and 10 GPa, 49 K for $A_xFe_{2-y}Se_2$ [20], respectively. It seems that the critical pressure $P_c$ depends on the bonding strength between the FeSe and the intercalation layers; the weakest chemical bonding in $Li_x(NH_3)_yFe_2Se_2$ gives rise to the lowest $P_c$. On the other hand, the maximum $T_c$ achievable in the SC-II phase seems to be proportional to the initial $T_c$, or the electron doping level, at ambient pressure. The observed maximum $T_c^{onset} \approx 55$ K in $Li_x(NH_3)_yFe_2Se_2$ is very close to the highest $T_c$ achieved in the FeAs-based materials [27].

In order to further characterize the SC-II phase, we tentatively probe the information about Fermi surface under high pressure by measuring the Hall effect in the normal state just above $T_c$. Fig. 4(a) shows the field dependence of in-plane Hall resistivity $\rho_{xy}(H)$ at 50 K for different pressures up to 9.5 GPa. All $\rho_{xy}(H)$ curves exhibit a linear behavior with a negative slope, which confirms that the electron-type charge carriers still dominate the transport properties in the SC-II phase. In addition, the slope of the $\rho_{xy}(H)$ curves decreases gradually with pressure. To quantify this change, we obtained the Hall coefficient $R_H \equiv d\rho_{xy}/dH$ from the linear fitting to the $\rho_{xy}(H)$ curves and plotted the pressure dependence of $R_H$ at 50 K in Fig. 4(b). As can be seen, the magnitude of $|R_H|$ decreases quickly and tends to level off above 6 GPa, in line with the variation of $T_c^{onset}(P)$ of the SC-II phase shown in Fig. 3. By assuming a simple one-band contribution, the electron-type carrier concentration can be estimated as $n_e = -1/(R_H \cdot e)$. As shown in Fig. 4(b), $n_e$ takes a value of ~ 0.4 ×10$^{27}$ m$^{-3}$ at 2 GPa, and experiences a fourfold enhancements to ~1.6 ×10$^{27}$ m$^{-3}$ at 7 GPa before leveling off, tracking nicely the variation of $T_c^{onset}(P)$. It should be noted that the carrier density at 7 GPa is slightly higher than that of 1.3 ×10$^{27}$ m$^{-3}$ at ambient pressure [24]. We have observed similar concomitant enhancement of $n_e$ and $T_c$ in the pressure-induced SC-II phase of (Li$_{1-x}$Fe$_x$)OHFeSe [23], thus implying a common mechanism controlling the $T_c$ of the SC-II phase in these HED FeSe-derived materials. As illustrated in Fig. 4(c), the $T_c^{onset}$ of the SC-II phases for both compounds are indeed found to scale linearly with the inverse Hall coefficient $R_H/R_H(P_c)$ or the electron charge density $n_e$, similar to the well-known Uemura's law [28]. In addition, these two curves are nearly parallel with each other, further elaborating a common origin for the pressure-induced SC-II phase.

As mentioned above, the electron doping plays an essential role to raise the $T_c$ of bulk FeSe, giving rise a variety of HED FeSe-derived bulk materials with an optimal $T_c$ reaching about 46 K at ambient pressure. The observed antiferromagnetic insulating behavior in the overdoped regime suggests the presence of some threshold for band filling to approach the Mott transition [18,29]. This will unavoidably set an upper limit of $T_c$ for these HED FeSe-based bulk materials at ambient pressure. If the band structure or Fermi surface topology keeps intact under pressure, the application of high pressure usually broadens the bandwidth, reduces the effective density of states at Fermi level, and then leads to a gradual reduction of $T_c$ as seen in the SC-I phase.

Since the structural transition has been excluded around $P_c$ in (Li,Fe)OHFeSe [23] and is unlikely to occur at such a low pressure of 2 GPa in Li$_x$(NH$_3$)$_y$Fe$_2$Se$_2$ [30], the sudden reversal of $T_c(P)$ above $P_c$ and the emergence of SC-II phase should be ascribed to an electronic origin, presumably associated with a Fermi surface reconstruction. The observed higher carrier concentration in the SC-II phase than that of SC-I phase, *e.g.* 1.6 ×10$^{27}$ m$^{-3}$ at 6 GPa versus 1.3×10$^{27}$ m$^{-3}$ at ambient pressure [24], indicates that the band structure of the SC-II phase would allow for more band filling before reaching the Mott transition. It is likely that the Fermi surface volume is enlarged above $P_c$. According to a recent ARPES study on FeSe films by Phan *et al.* [31], a compression strain realized in FeSe/CaF$_2$ will enlarge significantly both the hole and electron Fermi surfaces in comparison with the strain-free FeSe. Similarly, in these HED FeSe-based materials, there may also exist a critical pressure $P_c$ above which the compression on the FeSe planes can result in a sudden Fermi surface reconstruction or Lifshitz transition leading to a larger Fermi surface volume. The Se-Fe-Se angles and the anion height that can be tuned by pressure should be the key factors governing such a transition [32]. In addition, the observed concomitant enhancement of $T_c$ and $n_e$ in the SC-II phase suggests that the Fermi surface topology in the SC-II phase allows for above $P_c$ a gradual recovery of the density of states that has been reduced in the SC-I phase.

Nevertheless, it should be noted that the mechanism of the high $T_c$ superconductivity in the HED FeSe-derived materials is different from that of bulk FeSe, which exhibits enhanced $T_c$ to about 38.5 K under pressure with the dominant *hole* carriers [33]. Theoretical investigations in the $A_xFe_{2-y}Se_2$ systems have proposed that these two superconducting phases may have different pairing symmetries associated with a renormalization of Fermi surface topology [34]. Although much endeavor is needed to figure out the underlying mechanisms, our present work together with those previous high-pressure studies demonstrate that these HED FeSe-derived materials will universally emerge a pressure-induced SC-II phase with the maximum $T_c$ about 10 K higher than that of the SC-I phase [20,22,23]. This offers an alternative route to further raise $T_c$ of these HED FeSe-derived materials.

## Conclusion

In summary, we have performed the magnetotransport measurements on $Li_x(NH_3)_yFe_2Se_2$ single crystals under hydrostatic pressure up to 12 GPa, and constructed the *T-P* phase diagram featured by the emergence of a second high-$T_c$ superconducting phase above $P_c \approx 2$ GPa. We have achieved the highest $T_c^{onset} \approx 55$ K above 10 GPa among the FeSe-based **bulk** materials. In addition, we obtained a nearly parallel scaling behavior between $T_c^{onset}$ and the inverse Hall coefficient for the SC-II phase of both $Li_x(NH_3)_yFe_2Se_2$ and (Li,Fe)OHFeSe. Our present work thus demonstrates a unique way of high pressure to further raising $T_c$ of these HED FeSe-based materials by increasing the effective charge carrier concentration via a plausible Fermi surface reconstruction at $P_c$.

## Acknowledgements

This work is supported by the National Science Foundation of China (Grant Nos. 11574377, 11574394, 11774423), the National Basic Research Program of China (Grant No. 2014CB921500), the National Key R&D Program of China (Grant No. 2016YFA0300504), the Strategic Priority Research Program and Key Research Program of Frontier Sciences of the Chinese Academy of Sciences (Grant Nos. XDB07020100, QYZDB-SSW-SLH013), and the Fundamental Research Funds for the Central Universities, and the Research Funds of Renmin University of China (Grant Nos. 15XNLF06, 15XNLQ07).

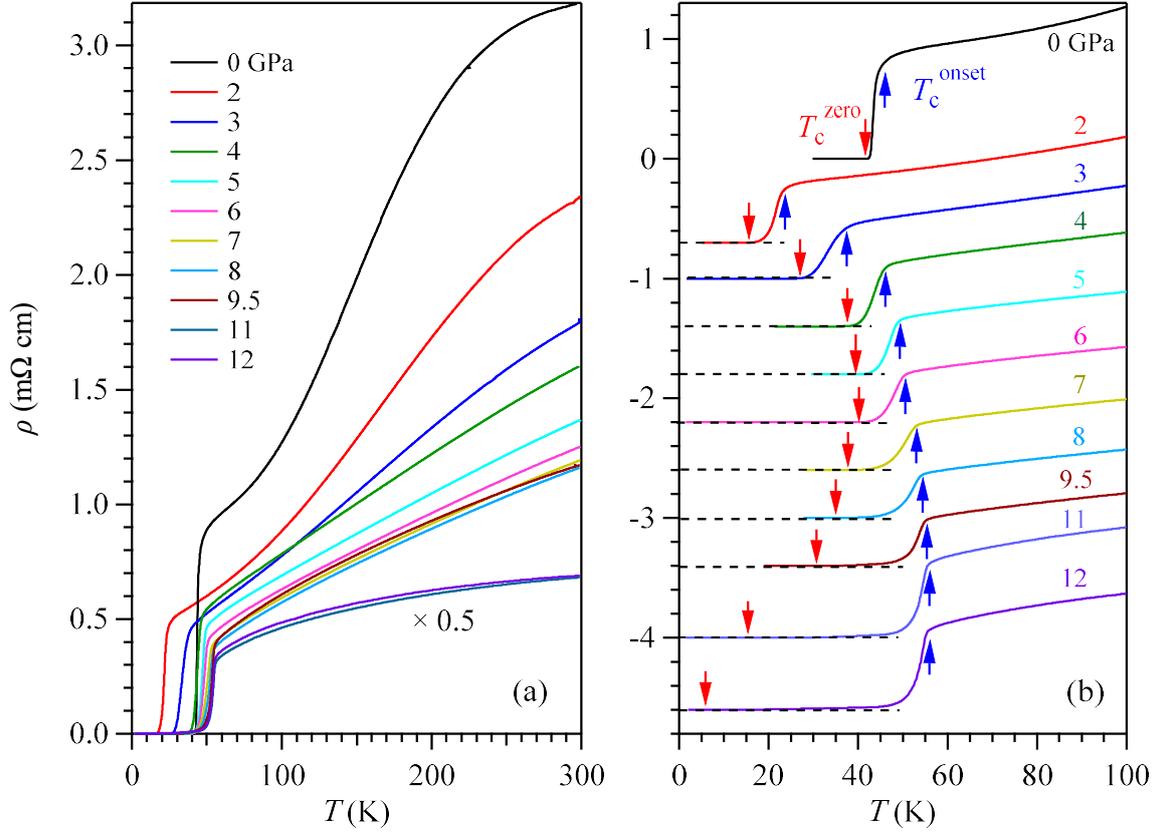

Fig. 1 High-pressure resistivity $\rho(T)$ for Li$_x$(NH$_3$)$_y$Fe$_2$Se$_2$ single crystal. (a) $\rho(T)$ curves in the whole temperature range illustrating the overall behaviors under pressure up to 12 GPa. At 11 and 12 GPa, the $\rho(T)$ curves were scaled by a factor of 0.5. (b) $\rho(T)$ curves below 100 K illustrating the variation of the superconducting transition temperatures with pressure. Except for data at 0 GPa, all other curves in (b) have been vertically shifted for clarity. The onset $T_c^{onset}$ (up-pointing arrow) was determined as the temperature where $\rho(T)$ curves above and below intersect with each other, while the $T_c^{zero}$ (down-pointing arrow) was determined as the zero-resistivity temperature.

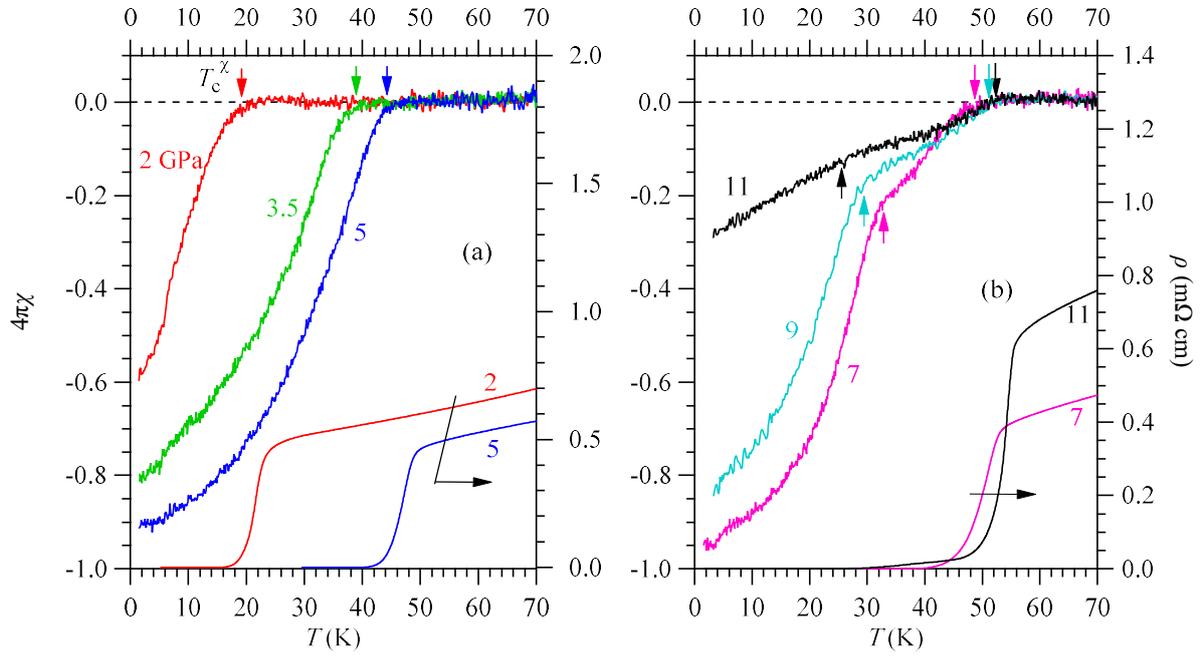

Fig. 2 Ac magnetic susceptibility $4\pi\chi(T)$ curves and resistivity $\rho(T)$ curves measured under different pressures up to 11 GPa. (a) The superconducting diamagnetic signal $4\pi\chi(T)$ and $\rho(T)$ below 5 GPa, the $T_c^\chi$ is in agreement with zero resistivity. (b) The superconducting diamagnetic signal $4\pi\chi(T)$ and $\rho(T)$ up to 11 GPa. Two transitions are marked by arrows, and are in agreement with zero resistivity and the onset of the superconductivity.

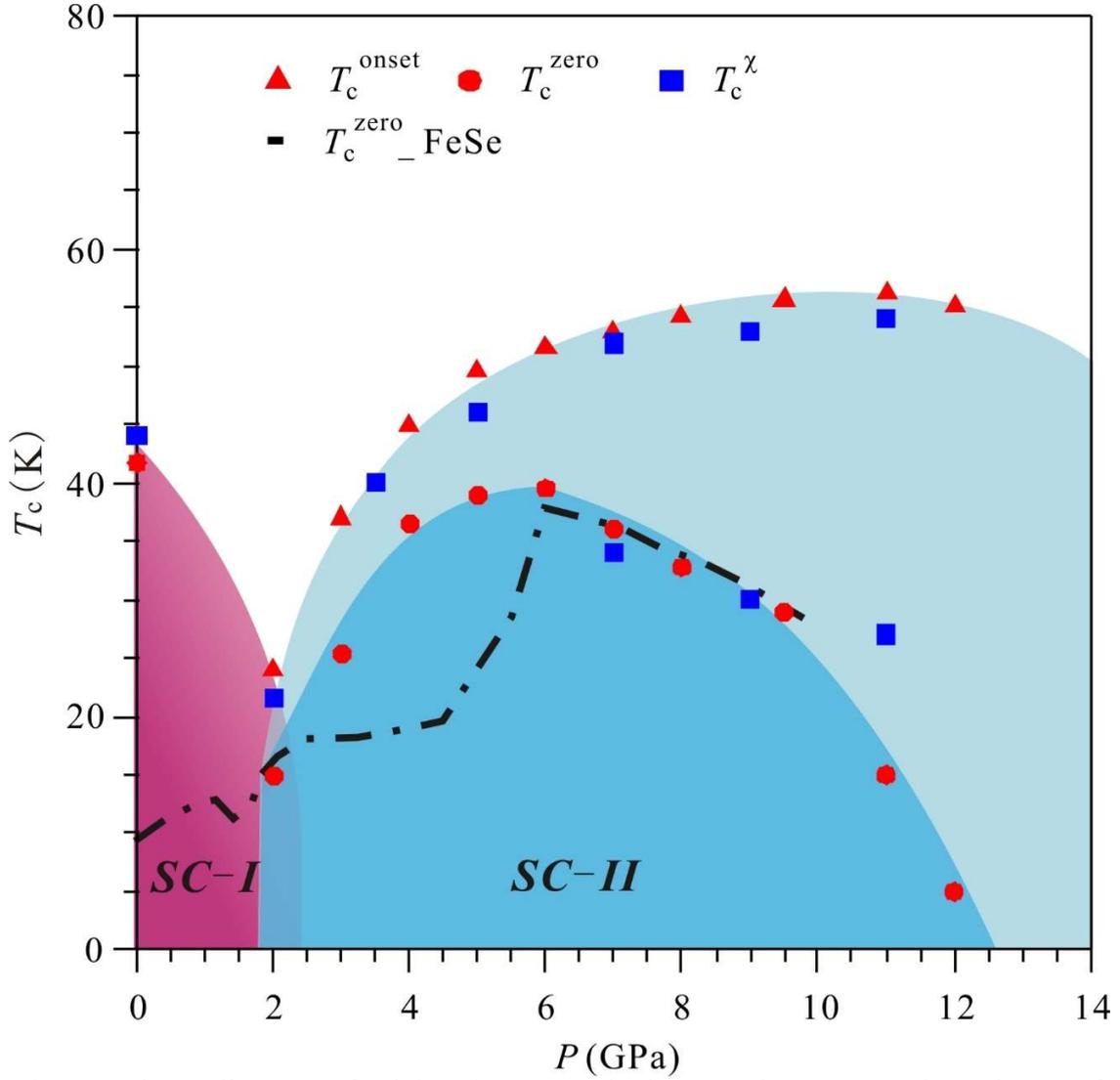

Fig. 3 *T-P* phase diagram of $Li_x(NH_3)_yFe_2Se_2$ single crystal. Pressure dependence of the superconducting transition temperatures $T_c$s up to 12 GPa. The values of $T_c^{onset}$, $T_c^{zero}$, and $T_c^{\chi}$ are determined from the high-pressure resistivity and ac magnetic susceptibility shown in Figs. 1 and 2. The $T_c^{zero}$ of FeSe single crystal is taken from Ref. [26].

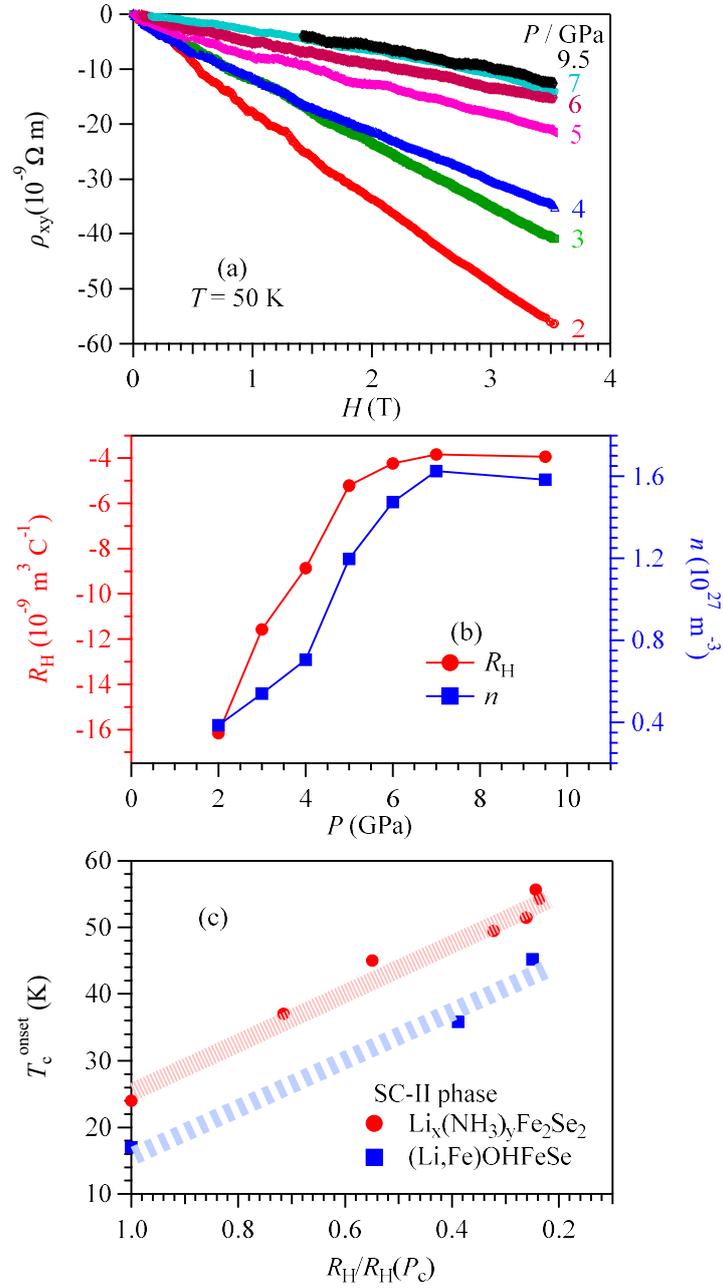

Fig. 4 (a) The Hall resistivity $\rho_{xy}$ at the normal state just above $T_c$ under various pressures. (b) The Hall coefficient $R_H$ and the carrier density $n_e$ are determined from the field derivative of $\rho_{xy}$, $R_H \equiv d\rho_{xy}/dH$ and $n_e = -1/(R_H \cdot e)$, at each pressure. (c) $T_c^{onset}$ as the dependence of the $R_H/R_H(P_c)$ in the SC-II phase of the $Li_x(NH_3)_yFe_2Se_2$ and (Li,Fe)OHFeSe single crystal [23].